

            Self-Duality and  Supersymmetry

                     E.Sezgin

                USE PLAIN TEX


\catcode`@=11

\def\oneandahalfspace{\baselineskip=1.15\normalbaselineskip plus 1pt
\lineskip=2pt\lineskiplimit=1pt}

\def\nofirstpagenoten{\nopagenumbers\footline={\ifnum\pageno>1\tenrm
\hss\folio\hss\fi}}
\def\nofirstpagenotwelve{\nopagenumbers\footline={\ifnum\pageno>1\twelverm
\hss\folio\hss\fi}}
\def\leaderfill{\leaders\hbox to 1em{\hss.\hss}\hfill}
\def\ft#1#2{{\textstyle{{#1}\over{#2}}}}
\def\frac#1/#2{\leavevmode\kern.1em
\raise.5ex\hbox{\the\scriptfont0 #1}\kern-.1em/\kern-.15em
\lower.25ex\hbox{\the\scriptfont0 #2}}
\def\sfrac#1/#2{\leavevmode\kern.1em
\raise.5ex\hbox{\the\scriptscriptfont0 #1}\kern-.1em/\kern-.15em
\lower.25ex\hbox{\the\scriptscriptfont0 #2}}

\parindent=20pt
\def\narrow{\advance\leftskip by 40pt \advance\rightskip by 40pt}

\def\nonarrower{\advance\leftskip by -40pt\advance\rightskip by -40pt}

\def\boxit#1{\vbox{\hrule\hbox{\vrule\kern3pt
        \vbox{\kern3pt#1\kern3pt}\kern3pt\vrule}\hrule}}

\def\gtorder{\mathrel{\raise.3ex\hbox{$>$}\mkern-14mu
             \lower0.6ex\hbox{$\sim$}}}
\def\ltorder{\mathrel{\raise.3ex\hbox{$<$}|mkern-14mu
             \lower0.6ex\hbox{\sim$}}}
\def\dalemb#1#2{{\vbox{\hrule height .#2pt
        \hbox{\vrule width.#2pt height#1pt \kern#1pt
                \vrule width.#2pt}
        \hrule height.#2pt}}}
\def\square{\mathord{\dalemb{4.9}{5}\hbox{\hskip1pt}}}

\font\fourteentt=cmtt10 scaled \magstep2
\font\fourteenbf=cmbx12 scaled \magstep1
\font\fourteenrm=cmr12 scaled \magstep1
\font\fourteeni=cmmi12 scaled \magstep1
\font\fourteenssr=cmss12 scaled \magstep1
\font\fourteenmbi=cmmib10 scaled \magstep2
\font\fourteensy=cmsy10 scaled \magstep2
\font\fourteensl=cmsl12 scaled \magstep1
\font\fourteenex=cmex10 scaled \magstep2
\font\fourteenit=cmti12 scaled \magstep1
\font\twelvett=cmtt10 scaled \magstep1 \font\twelvebf=cmbx12
\font\twelverm=cmr12  \font\twelvei=cmmi12
\font\twelvessr=cmss12 \font\twelvembi=cmmib10 scaled \magstep1
\font\twelvesy=cmsy10 scaled \magstep1
\font\twelvesl=cmsl12 \font\twelveex=cmex10 scaled \magstep1
\font\twelveit=cmti12
\font\tenssr=cmss10 \font\tenmbi=cmmib10
 
 \font\ninebf=cmbx9
\font\ninerm=cmr9  \font\ninei=cmmi9
\font\ninesy=cmsy9 \font\ninessr=cmss9
\font\ninembi=cmmib10 scaled 900
\font\eightit=cmti8 \font\eightsl=cmsl8
\font\eighttt=cmtt8 \font\eightbf=cmbx8
\font\eightrm=cmr8  \font\eighti=cmmi8
\font\eightsy=cmsy8 \font\eightex=cmex10 scaled 800
\font\eightssr=cmss8 \font\eightmbi=cmmib10 scaled 800
 
\font\sevenbf=cmbx7 \font\sevenrm=cmr7 \font\seveni=cmmi7
\font\sevensy=cmsy7 
\font\sevenssr=cmss9 scaled 778 \font\sevenmbi=cmmib10 scaled 700
 
 \font\sixbf=cmbx7 scaled 875
\font\sixrm=cmr6  \font\sixi=cmmi6
\font\sixsy=cmsy6 \font\sixssr=cmss8 scaled 750
\font\sixmbi=cmmib10 scaled 600
\font\fivessr=cmss8 scaled 625  \font\fivembi=cmmib10 scaled 500

\newskip\ttglue
\newfam\ssrfam
\newfam\mbifam

\mathchardef\alpha="710B
\mathchardef\beta="710C
\mathchardef\gamma="710D
\mathchardef\delta="710E
\mathchardef\epsilon="710F
\mathchardef\zeta="7110
\mathchardef\eta="7111
\mathchardef\theta="7112
\mathchardef\iota="7113
\mathchardef\kappa="7114
\mathchardef\lambda="7115
\mathchardef\mu="7116
\mathchardef\nu="7117
\mathchardef\xi="7118
\mathchardef\pi="7119
\mathchardef\rho="711A
\mathchardef\sigma="711B
\mathchardef\tau="711C
\mathchardef\upsilon="711D
\mathchardef\phi="711E
\mathchardef\chi="711F
\mathchardef\psi="7120
\mathchardef\omega="7121
\mathchardef\varepsilon="7122
\mathchardef\vartheta="7123
\mathchardef\varpi="7124
\mathchardef\varrho="7125
\mathchardef\varsigma="7126
\mathchardef\varphi="7127
\mathchardef\partial="7140

\def\fourteenpoint{\def\rm{\fam0\fourteenrm}
\textfont0=\fourteenrm \scriptfont0=\tenrm \scriptscriptfont0=\sevenrm
\textfont1=\fourteeni \scriptfont1=\teni \scriptscriptfont1=\seveni
\textfont2=\fourteensy \scriptfont2=\tensy \scriptscriptfont2=\sevensy
\textfont3=\fourteenex \scriptfont3=\fourteenex \scriptscriptfont3=\fourteenex
\def\it{\fam\itfam\fourteenit} \textfont\itfam=\fourteenit
\def\sl{\fam\slfam\fourteensl} \textfont\slfam=\fourteensl
\def\bf{\fam\bffam\fourteenbf} \textfont\bffam=\fourteenbf
\scriptfont\bffam=\tenbf \scriptscriptfont\bffam=\sevenbf
\def\tt{\fam\ttfam\fourteentt} \textfont\ttfam=\fourteentt
\def\ssr{\fam\ssrfam\fourteenssr} \textfont\ssrfam=\fourteenssr
\scriptfont\ssrfam=\tenmbi \scriptscriptfont\ssrfam=\sevenmbi
\def\mbi{\fam\mbifam\fourteenmbi} \textfont\mbifam=\fourteenmbi
\scriptfont\mbifam=\tenmbi \scriptscriptfont\mbifam=\sevenmbi
\tt \ttglue=.5em plus .25em minus .15em
\normalbaselineskip=16pt
\bigskipamount=16pt plus5pt minus5pt
\medskipamount=8pt plus3pt minus3pt
\smallskipamount=4pt plus1pt minus1pt
\abovedisplayskip=16pt plus 4pt minus 12pt
\belowdisplayskip=16pt plus 4pt minus 12pt
\abovedisplayshortskip=0pt plus 4pt
\belowdisplayshortskip=9pt plus 4pt minus 6pt
\parskip=5pt plus 1.5pt
\twelvefoot
\setbox\strutbox=\hbox{\vrule height12pt depth5pt width0pt}
\let\sc=\tenrm
\let\big=\fourteenbig \normalbaselines\rm}
\def\fourteenbig#1{{\hbox{$\left#1\vbox to12pt{}\right.\n@space$}}
\def\square{\mathord{\dalemb{6.8}{7}\hbox{\hskip1pt}}}}

\def\twelvepoint{\def\rm{\fam0\twelverm}
\textfont0=\twelverm \scriptfont0=\ninerm \scriptscriptfont0=\sevenrm
\textfont1=\twelvei \scriptfont1=\ninei \scriptscriptfont1=\seveni
\textfont2=\twelvesy \scriptfont2=\ninesy \scriptscriptfont2=\sevensy
\textfont3=\twelveex \scriptfont3=\twelveex \scriptscriptfont3=\twelveex
\def\it{\fam\itfam\twelveit} \textfont\itfam=\twelveit
\def\sl{\fam\slfam\twelvesl} \textfont\slfam=\twelvesl
\def\bf{\fam\bffam\twelvebf} \textfont\bffam=\twelvebf
\scriptfont\bffam=\ninebf \scriptscriptfont\bffam=\sevenbf
\def\tt{\fam\ttfam\twelvett} \textfont\ttfam=\twelvett
\def\ssr{\fam\ssrfam\twelvessr} \textfont\ssrfam=\twelvessr
\scriptfont\ssrfam=\ninessr \scriptscriptfont\ssrfam=\sevenssr
\def\mbi{\fam\mbifam\twelvembi} \textfont\mbifam=\twelvembi
\scriptfont\mbifam=\ninembi \scriptscriptfont\mbifam=\sevenmbi
\tt \ttglue=.5em plus .25em minus .15em
\normalbaselineskip=14pt
\bigskipamount=14pt plus4pt minus4pt
\medskipamount=7pt plus2pt minus2pt
\abovedisplayskip=14pt plus 3pt minus 10pt
\belowdisplayskip=14pt plus 3pt minus 10pt
\abovedisplayshortskip=0pt plus 3pt
\belowdisplayshortskip=8pt plus 3pt minus 5pt
\parskip=3pt plus 1.5pt
\tenfoot
\setbox\strutbox=\hbox{\vrule height10pt depth4pt width0pt}
\let\sc=\ninerm
\let\big=\twelvebig \normalbaselines\rm}
\def\twelvebig#1{{\hbox{$\left#1\vbox to10pt{}\right.\n@space$}}
\def\square{\mathord{\dalemb{5.9}{6}\hbox{\hskip1pt}}}}

\def\tenpoint{\def\rm{\fam0\tenrm}
\textfont0=\tenrm \scriptfont0=\sevenrm \scriptscriptfont0=\fiverm
\textfont1=\teni \scriptfont1=\seveni \scriptscriptfont1=\fivei
\textfont2=\tensy \scriptfont2=\sevensy \scriptscriptfont2=\fivesy
\textfont3=\tenex \scriptfont3=\tenex \scriptscriptfont3=\tenex
\def\it{\fam\itfam\tenit} \textfont\itfam=\tenit
\def\sl{\fam\slfam\tensl} \textfont\slfam=\tensl
\def\bf{\fam\bffam\tenbf} \textfont\bffam=\tenbf
\scriptfont\bffam=\sevenbf \scriptscriptfont\bffam=\fivebf
\def\tt{\fam\ttfam\tentt} \textfont\ttfam=\tentt
\def\ssr{\fam\ssrfam\tenssr} \textfont\ssrfam=\tenssr
\scriptfont\ssrfam=\sevenssr \scriptscriptfont\ssrfam=\fivessr
\def\mbi{\fam\mbifam\tenmbi} \textfont\mbifam=\tenmbi
\scriptfont\mbifam=\sevenmbi \scriptscriptfont\mbifam=\fivembi
\tt \ttglue=.5em plus .25em minus .15em
\normalbaselineskip=12pt
\bigskipamount=12pt plus4pt minus4pt
\medskipamount=6pt plus2pt minus2pt
\abovedisplayskip=12pt plus 3pt minus 9pt
\belowdisplayskip=12pt plus 3pt minus 9pt
\abovedisplayshortskip=0pt plus 3pt
\belowdisplayshortskip=7pt plus 3pt minus 4pt
\parskip=0.0pt plus 1.0pt
\eightfoot
\setbox\strutbox=\hbox{\vrule height8.5pt depth3.5pt width0pt}
\let\sc=\eightrm
\let\big=\tenbig \normalbaselines\rm}
\def\tenbig#1{{\hbox{$\left#1\vbox to8.5pt{}\right.\n@space$}}
\def\square{\mathord{\dalemb{4.9}{5}\hbox{\hskip1pt}}}}

\def\eightpoint{\def\rm{\fam0\eightrm}
\textfont0=\eightrm \scriptfont0=\sixrm \scriptscriptfont0=\fiverm
\textfont1=\eighti \scriptfont1=\sixi \scriptscriptfont1=\fivei
\textfont2=\eightsy \scriptfont2=\sixsy \scriptscriptfont2=\fivesy
\textfont3=\eightex \scriptfont3=\eightex \scriptscriptfont3=\eightex
\def\it{\fam\itfam\eightit} \textfont\itfam=\eightit
\def\sl{\fam\slfam\eightsl} \textfont\slfam=\eightsl
\def\bf{\fam\bffam\eightbf} \textfont\bffam=\eightbf
\scriptfont\bffam=\sixbf \scriptscriptfont\bffam=\fivebf
\def\tt{\fam\ttfam\eighttt} \textfont\ttfam=\eighttt
\def\ssr{\fam\ssrfam\eightssr} \textfont\ssrfam=\eightssr
\scriptfont\ssrfam=\sixssr \scriptscriptfont\ssrfam=\fivessr
\def\mbi{\fam\mbifam\eightmbi} \textfont\mbifam=\eightmbi
\scriptfont\mbifam=\sixmbi \scriptscriptfont\mbifam=\fivembi
\tt \ttglue=.5em plus .25em minus .15em
\normalbaselineskip=9pt
\bigskipamount=9pt plus3pt minus3pt
\medskipamount=5pt plus2pt minus2pt
\abovedisplayskip=9pt plus 3pt minus 9pt
\belowdisplayskip=9pt plus 3pt minus 9pt
\abovedisplayshortskip=0pt plus 3pt
\belowdisplayshortskip=5pt plus 3pt minus 4pt
\parskip=0.0pt plus 1.0pt
\setbox\strutbox=\hbox{\vrule height8.5pt depth3.5pt width0pt}
\let\sc=\sixrm
\let\big=\eightbig \normalbaselines\rm}
\def\eightbig#1{{\hbox{$\left#1\vbox to6.5pt{}\right.\n@space$}}
\def\square{\mathord{\dalemb{3.9}{4}\hbox{\hskip1pt}}}}

\def\vfootnote#1{\insert\footins\bgroup\footsuite
    \interlinepenalty=\interfootnotelinepenalty
    \splittopskip=\ht\strutbox
    \splitmaxdepth=\dp\strutbox \floatingpenalty=20000
    \leftskip=0pt \rightskip=0pt \spaceskip=0pt \xspaceskip=0pt
    \textindent{#1}\footstrut\futurelet\next\fo@t}
\def\hangfootnote#1{\edef\@sf{\spacefactor\the\spacefactor}#1\@sf
    \insert\footins\bgroup\footsuite
    \let\par=\endgraf
    \interlinepenalty=\interfootnotelinepenalty
    \splittopskip=\ht\strutbox
    \splitmaxdepth=\dp\strutbox \floatingpenalty=20000
    \leftskip=0pt \rightskip=0pt \spaceskip=0pt \xspaceskip=0pt
    \smallskip\item{#1}\bgroup\strut\aftergroup\@foot\let\next}
\def\footsuite{}
\def\twelvefoot{\def\footsuite{\twelvepoint}}
\def\tenfoot{\def\footsuite{\tenpoint}}
\def\eightfoot{\def\footsuite{\eightpoint}}
\catcode`@=12


\def\ft#1#2{{\textstyle{{#1}\over{#2}}}}

\def\semidirprod{\rlap{\ss C}\raise1pt\hbox{$\mkern.75mu\times$}}
\def\i{{\rm i}}

\def\for{\lower6pt\hbox{$\Big|$}}

\nofirstpagenotwelve
\twelvepoint
\def\fish{\kern-.25em{\phantom{abcde}\over \phantom{abcde}}\kern-.25em}

\oneandahalfspace
\overfullrule=0pt

\line{\hfil CTP-TAMU-85/92}
\line{\hfil hepth@xxx/9212092}
\line{\hfil December 1992}
\vskip 2truecm
\centerline{\bf  Self-Duality and Supersymmetry}
\vskip 2truecm
 \centerline{ E. Sezgin {\footnote{$^*$} { Supported in part by NSF, under
grant
PHY-9106593.\hfill\break
 \phantom\quad\quad  Contribution to the Proceedings of the Trieste Summer
Workshop on
Superstrings and Related Topics, \hfill\break
\phantom\quad\quad  Trieste, July 1992.} } }
\vskip 1truecm
 \centerline{\it Center for Theoretical Physics, Texas A\&M University,}
\centerline{\it College Station, Texas 77843-4242, USA.}  \vskip .4truein
\vskip 2truecm
\centerline{\bf ABSTRACT}
\vskip .5truecm
 {\tenfoot
Chiral/self-dual restrictions of various super Yang-Mills  and
supergravity theories in (2,2) dimensions are described.  These include  the
N=1 supergravity
with a cosmological term and the N=1 new minimal supergravity theory.  In the
latter case,
a self-duality condition on a  torsionful Riemann curvature is possible, and it
implies the
equations of motion that follow from an $R^2$ type  supergravity action.
}
\vfill\eject
\noindent{\bf 1.}\   Self-dual Yang-Mills and self-dual gravity equations in
(2,2) dimensions are known to arise  in open and closed N=2 string theories
[1],
i.e. string
theories  with {\it local}  N=2  world-sheet supersymmetry.  Recently, it has
been shown that
N=2 open string  actually describes a self-dual  {\it super} Yang-Mills theory,
while the N=2
closed string  describes  self-dual N=8 supergravity theory and the heterotic
N=2 closed string
describes  {\it gauged} N=8 supergravity  theory in (2,2) dimensions [2].
These
interesting
developments motivate further  study of various self-dual supersymmetric field
 theories in
(2,2) dimensions.

Another motivation for studying the self-dual systems in (2,2) dimensions is
their connection
with a large class of integrable systems in (1,1) dimensions [3].  For
example,
different types of
dimensional
reductions of the (2,2) dimensional self-dual Yang-Mills theory yield
a large class
of  integrable
systems in (1,1) dimensions including  the KdV equation, the
non-linear
Schrodinger
equations and Toda equations.  One expects that even a larger class of
integrable
models are e
mbedded in self-dual supersymmetric field theories.  In fact, it has
recently been
shown that super KdV and super-Toda equations do arise from  dimensional
 reductions of N=2
supersymmetric self-dual  Yang-Mills system in (2,2) dimensions [4].

In this note, we shall describe the algebraic structure of various self-dual
supersymmetric
field theories in (2,2) dimensions.  We first describe their underlying
superalgebras.
Next, we
discuss the
chiral/self-dual restrictions of N=1,2,4 super Yang-Mills theories.
Finally, we
describe the chiral/self-dual restrictions of (i)  N=1 supergravity
(ii) new minimal N=1
supergravity and (iii) N=1 supergravity with a cosmological constant in (2,2)
 dimensions. In the
second case,  we find a  self-duality condition on a torsionful Riemann
curvature which
implies the equations of motion that follow from an $R^2$ type action [5].
In the last case,
surprisingly,  it turns out that (a) the self-duality conditions on the
Riemann tensor and
the gravitino curvature are precisely those of self-dual  N=1
supergravity without a cosmological constant, and (b) in the chiral
truncation of the
supersymmetry transformation rules  of anti de Sitter supergravity,
the cosmological
constant parameter survives. Thus, the usual self-dual supergravity
equations seem to
 be invariant under  a one parameter extension of the super-Poincar\'e
group which is a
particular contraction of the anti de Sitter Poincar\'e algebra in
(2,2) dimensions.  A similar
phenomenon  has been descibred in the context of gauged N=8
supergravity in [2].

  \bigskip
\noindent{\bf 2.}\  Let us  consider  first the relevant super
 Poincar\'e algebras.  A
characteristic feature of the super Poincar\'e algebra in (2,2) dimensions
is that  the
supercharges are  independent real Majorana-Weyl spinors [6].  Various
supermultiplets in (3,1)
dimensions can easily be Wick rotated to  supermultiplets in (2,2) dimensions.
Of particular
interest is the Yang-Mills supermultiplet consisting of the Yang-Mills gauge
field and its
fermionic partner.
 We recall  that this   supermultiplet also exists in (5,1), where the
supercharges  are
pseudo-real Majorana-Weyl, and in (9,1) dimensions where the supercharges
are real
Majorana-Weyl.  We can Wick rotate the first one to (3,3) dimensions,
and the latter
to (5,5) dimensions. In both cases, there exist real Majorana-Weyl spinors.
Thus,
it is  useful to consider the  (2,2), (3,3), (5,5) and (9,1) dimensions,
where  real Majorana-Weyl spinors and simple Yang-Mills supermultiplets exist.

In (2,2) dimensions, we adopt the following conventions.
 The spacetime signature is $(--++)$, the charge conjugation
matrix is  $C=\gamma_{12}$,
$C^T=-C$, $(\gamma_{\mu\nu}C)^T=\gamma_{\mu\nu}C$,
$\gamma_5=\gamma_{1234}$,
$\gamma_5^T=\gamma_5$, $[C,\gamma_5]=0$,
$\gamma_\mu^*=\eta \gamma_\mu$, and
$(\gamma_\mu C)^T=-\eta\gamma_\mu C$, where
$\eta=\pm 1$.  Without loss of generality, we shall set $\eta=1$.
The extended Poincar\'e
algebra consists of Poincar\'e generators and supercharges
$Q_\pm^i\ ,i=1,...,N$   which
obey the anticommutation rules
$$
 \eqalign{
\{Q_+^i, Q_-^j\} &= \ft12
(1+\gamma_5)\gamma^\mu C P_\mu\lambda^{ij} \ , \cr
\{Q^i_\pm, Q^j_\pm\} &= 0\ , \cr
}\eqno(1)
$$
where $\lambda_{ij}$  can be chosen to be either the $SO(n, N-n)$
invariant symmetric tensor
$\eta_{ij}$ or the $Sp(n,N-n)$ invariant antisymmetric tensor $\Omega_{ij}$.
These are the
automorphism groups of the algebra, and the supercharges transform in
their defining
representations.

In (3,3) dimensions,  we adopt the following conventions:
 The spacetime signature is $(---+++)$,  the charge conjugation matrix is
$C=\gamma_{123}$,
$(\gamma_\mu C)^T=-(\gamma_\mu C)$,
$(\gamma_{\mu\nu\rho}C)^T=\gamma_{\mu\nu\rho}C$,
$\gamma_7=\gamma_{123456}$,
$\gamma_7^T=\gamma_7$, $\{C,\gamma_7\}=0$,
$\gamma_\mu^*=\eta \gamma_\mu$, $C^T=-\eta C$ and
$(\gamma_{\mu\nu}C)^T=-\eta\gamma_{\mu\nu}C$,  where $\eta=\pm 1$.
Without loss of
generality, we shall set $\eta=1$.  The  supercharges  $Q_+^i\ ,i=1,...,M$
and $Q_-^a,\ a=1,...,N$
obey the anticommutation rules
$$
\eqalign{
\{Q_+^i, Q_+^j\} &= \ft12 (1+\gamma_7)\gamma^\mu C P_\mu\Omega^{ij}
 \ ,\cr
\{ Q^i_+, Q^a_- \} &= 0\ , \cr
\{Q_+^a, Q_+^b\} &= \ft12 (1+\gamma_7)\gamma^\mu C P_\mu\Omega^{ab}
\ , \cr
}\eqno(2)
$$
where $\Omega^{ij}$ and $\Omega^{ab}$ are the invariant tensors of the
 automorphism
group $Sp(m, M-m)\times Sp(n,N-n)$.

Finally, in (5,5)
dimensions,  we adopt the following conventions.  The spacetime
signature
is $(-----+++++)$,  the charge conjugation matrix is $C=\gamma_{12345}$,
$(\gamma_\mu C)^T=(\gamma_\mu C)$,
$(\gamma_{\mu\nu\rho}C)^T=-\gamma_{\mu\nu\rho}C$,
$(\gamma_{\mu\nu\rho\lambda\tau}C)^T=\gamma_{\mu\nu\rho\lambda\tau}C$,
$\gamma_{11}=\gamma_{1...10}$,  $\gamma_{11}^2=1$,
$\gamma_{11}^T=\gamma_{11}$,
$\gamma_{11}^*=\gamma_{11}$, $\{C,\gamma_{11}\}=0$,
$\gamma_\mu^*=\eta \gamma_\mu$,
$C^T=-\eta C$, $(\gamma_{\mu\nu}C)^T=\eta\gamma_{\mu\nu}C$,  and
$(\gamma_{\mu\nu\rho\sigma}C)^T=-\eta\gamma_{\mu\nu\rho\sigma}C$,
where $\eta=\pm 1$.
Again, without loss of generality, we shall set $\eta=1$. The supercharges
$Q_+^i\ ,i=1,...,M$ and $Q_-^a,\ a=1,...,N$ obey the anticommutation rules
$$
\eqalign{
\{Q_+^i, Q_+^j\} &= \ft12 (1+\gamma_{11})\gamma^\mu C P_\mu\eta^{ij} \ , \cr
\{Q^i_+, Q^a_-\} &= 0\ , \cr
\{Q_+^a, Q_+^b\} &= \ft12 (1+\gamma_{11})\gamma^\mu C P_\mu\eta^{ab} \ ,\cr
}\eqno(3)
$$
where $\eta^{ij}$ and $\eta^{ab}$ are the invariant tensors of the
automorphism
group $SO(m,M-m)\times SO(n,N-n)$.  The properties of the Dirac matrices
 and the form of the
extended Poincar\'e algebra in $(9,1)$ dimensions is the same as in the
$(5,5)$ dimensions
described above.
\bigskip
\noindent{\bf 3.}\  We now turn to the description of self-dual super
Yang-Mills theories in
$(2,2)$ dimensions.  It can easily obtained by a Wick rotation of the
super Yang-Mills theory
in $(3,1)$ dimensions.  The model contains the gauge field $A_\mu$
 and its fermionic parners
$\lambda_\pm$, which are now independent real Majorana-Weyl spinors.
The self-dual
version of the theory is obtained by setting $\lambda_+=0$ [7].  The
 resulting transformation
rules are
 $$
\eqalign{
\delta A_\mu &= \ft12 \bar \epsilon_+\gamma_\mu \lambda_- \ ,\cr
\delta \lambda_- &= - \ft14 \gamma^{\mu\nu}F_{\mu\nu} \epsilon_-\ .\cr
}\eqno(4)
$$
Note that both supersymmetry parameters $\epsilon_\pm$
are present. The closure of these transformations  then requires
the conditions
 $$
\eqalign{
F_{\mu\nu} - \ft12 \epsilon_{\mu\nu\rho\sigma} F^{\rho\sigma}&=0,  \cr
 \gamma^\mu D_\mu \lambda_-  &=0\ ,  \cr
}
\eqno(5)
$$
where $D_\mu$ is the gauge covariant derivative.  These equations
transform into each other
under the supersymmetry transformations: $\delta F_{\mu\nu}^-=-\ft12
{\bar\epsilon}_+\gamma_{\mu\nu}\gamma^\rho D_\rho\lambda_-$ and
$\delta  \gamma^\mu D_\mu\lambda_-
=-\ft14 \gamma^\mu\gamma^{\nu\rho} D_\mu
F_{\nu\rho}^-\epsilon_-$, where $F_{\mu\nu}^-\equiv
F_{\mu\nu} - \ft12 \epsilon_{\mu\nu\rho\sigma} F^{\rho\sigma}$.

As to be expected, the N=1 self-dual super Yang-Mills equations (5)
are also invariant
the transformations of the superconformal group  $SL(4|1)$ whose
bosonic subgroup is
$SL(4,R) (\approx SO(3,3))\times SO(1,1)$. The superconformal
transformations are given
by (4) and
 $$
\eqalign{
\delta A_\mu &= -\ft12 \bar\eta_-\gamma^\nu\gamma_\mu\lambda_-
x_\nu + \xi^\lambda\partial_\lambda A_\mu\ ,\cr
\delta\lambda_- &=
-\ft14 \gamma^{\mu\nu}\gamma^\rho x_\rho F_{\mu\nu}
\eta_+ + \ft32\alpha\lambda_- - \beta\lambda_-
+\ft14\omega^{\mu\nu}\gamma_{\mu\nu}\lambda_-
+\xi^\mu\partial_\mu\lambda_- \ , \cr
}\eqno(6)
$$
where $\eta_\pm$ are the special conformal supersymmetry parameters,
$\alpha,\beta$ are the parameters of dilatations and SO(1,1)
transformations, respectively, and $\xi^\mu$ represent the
spacetime conformal transformations given by
$$
\xi^\mu(x)= a^\mu - \omega^{\mu\nu}x_\nu + \alpha x^\mu
+2x^\mu\eta \cdot x - \eta^\mu x^2 \ . \eqno(7)
$$
Here $a^\mu,\omega^{\mu\nu}$ and $\eta^\mu$ are the constant
parameters of translations, Lorentz rotations and conformal boosts,
respectively.
\bigskip
\noindent{\bf 4.}
The on-shell N=2 super Yang-Mills multiplet in (2,2) dimensions
contains the fields $(A_\mu, \lambda^i_\pm, S, \phi)$ where $i=1,2$
and $S,\phi$ are two real
scalars,  all taking their values in the adjoint representation of some
 Lie algebra.  It is
easiest to obtain the transformation rules for this multiplet from (3,3)
dimensions by
ordinary dimensional reduction. The  simple super Yang-Mills multiplet
transformation rules in
(3,3) dimensions are
$$
\eqalign{
\delta A_{\hat\mu} &= \ft12 \bar\eta^i\Gamma_{\hat \mu}
\chi^j\epsilon_{ij} \ , \qquad\qquad \hat\mu= 1,...,6 \cr
\delta\chi^i &= - \ft14 \Gamma^{\hat\mu\hat \nu} F_{\hat\mu\hat\nu}
\eta^i\ ,  \cr
}\eqno(8)
$$
where $\Gamma_7\chi =-\chi$ and  $\Gamma_7\eta =-\eta$.
Let us choose the Dirac matrices as $\Gamma^\mu=\gamma^\mu\times 1$,
$\mu=1,2,5,6$,
$\Gamma_3=\gamma_5\times i\sigma_2$, $\Gamma_4=
\gamma_5\times \sigma_1$.
 Performing a straigtforward dimensional
reduction  in which we take all the fields to be independent of the two
 internal coordinates
and defining $\phi\equiv \ft12 (A_3-A_4)$, and $S\equiv \ft12 (A_3+A_4)$,
from (8) we find
the transformation rules for the $N=2$ super Yang-Mills multiplet in
(2,2) dimensions:
$$
\eqalign{
\delta A_\mu &=
\ft12{\bar\eta}^i_+\gamma_\mu i\sigma_2 \chi_-^j\epsilon_{ij}
-\ft12{\bar\eta}^i_-\gamma_\mu i\sigma_2 \chi_+^j\epsilon_{ij}\ , \cr
\delta\phi &=-\ft12{\bar\eta}^i_-\chi_-^j\epsilon_{ij}\ ,\cr
 \delta S &=-\ft12{\bar\eta}^i_+\chi_+^j\epsilon_{ij}\ , \cr
\delta \chi_+^i &=-\ft14\gamma^{\mu\nu}F_{\mu\nu}\eta^i_+
-\gamma^\mu D_\mu\phi
\sigma_+ \eta^i_- +\gamma^\mu D_\mu S \sigma_-{\bar\eta}^i_-
- [S,\phi] \eta_+^i\ ,  \cr
\delta \chi_-^i &=-\ft14\gamma^{\mu\nu}F_{\mu\nu}\eta^i_-
+\gamma^\mu D_\mu\phi
\sigma_+ \eta^i_+ -\gamma^\mu D_\mu S \sigma_-{\bar\eta}^i_-
+[S,\phi] \eta_+^i\ ,  \cr
}\eqno(9)
$$
where $\gamma_5\chi_\pm^i=\pm\chi_\pm^i$ and $\sigma_\pm =
\ft12 (\sigma_1\pm
\i\sigma_2)$.  To obtain the self-dual version, we impose the conditions
 [8]
$$
   S=0\ ,\qquad \chi_+^i=0\ , \qquad  \sigma_+\chi_-^i=0\ ,\qquad
\sigma_+\eta_-^i=0\ ,\qquad  \sigma_-\eta_+^i=0\ .  \eqno(10)
$$
Using the notation,   $\sigma_-\chi_-^i \equiv \lambda_-^i$ and
$\sigma_+\eta_+^i \equiv \epsilon_+^i$, we then have
$$
\eqalign{
\delta A_\mu &=
\ft12{\bar\epsilon}^i_+\gamma_\mu\lambda_-^j\epsilon_{ij}\ ,  \cr
\delta\phi &=-\ft12{\bar\epsilon}^i_-\lambda_-^j\epsilon_{ij}\ , \cr
 \delta \lambda_-^i &=-\ft14\gamma^{\mu\nu}F_{\mu\nu}\epsilon^i_-
+\gamma^\mu D_\mu\phi
\epsilon^i_+ \ .   \cr
}\eqno(11)
$$
The
closure of the  algebra modulo a set of equations of motions that
transform into each
other under superymmetry  requires the following set of equations [8]
$$
\eqalign{
F_{\mu\nu} - \ft12 \epsilon_{\mu\nu\rho\sigma} F^{\rho\sigma}  &=0
\ ,\cr
\gamma^\mu D_\mu \lambda^i_-  &=0,  \cr
D_\mu D^\mu \phi-\ft18\epsilon_{ij}
[{\bar \lambda}^i_-, \lambda^j_-]  &=0\ , \cr
}\eqno(12)
$$
where the commutator is in the space of the Lie algebra generators.  The last
 equation is
easily seen
to follow from the supersymmetric variation of the second equation.
One application
of these equations has already been considered in [4] where it has been
an appropriate
dimensional reduction of these equations to (1,1) dimensions yields the
super KdV
equations.
\bigskip
\noindent {\bf  5.}\
We now consider the case of N=4 super Yang-Mills
theory.  While it can be formulated in (2,2) dimensions, it does not admit a
simple self-dual
truncation of the  kind discussed above.   The N=4 super Yang-Mills theory
in (2,2) dimensions
can be easily
 obtained by an ordinary dimensional reduction from the simple  super
Yang-Mills theory in
(5,5) dimensions, whose transformation rules are
$$
\eqalign{
\delta A_{\hat\mu} &= \ft12 {\bar\epsilon}\Gamma_{\hat \mu}\lambda\ ,
\qquad\qquad \hat\mu=1,...,10 \cr
\delta\lambda   &= - \ft14 \Gamma^{\hat\mu\hat \nu}
 F_{\hat\mu\hat\nu}\epsilon\ ,\cr
}\eqno(13)
$$
where $\Gamma_{11}\lambda=-\lambda$ and
$\Gamma_{11}\epsilon=-\epsilon$.  Let us choose the Dirac matrices
as $\Gamma_\mu=
\gamma_\mu\times 1$ $(\mu=1,2,9,10)$ and $\Gamma_i=
\gamma_5 \times \gamma_i$
 $(i=3,...,8)$.  As usual, setting $\partial_i=0$, we obtain the
N=4 super Yang-Mills
transformation rules
$$
\eqalign{
\delta A_\mu &= {\bar \epsilon}_+\gamma_\mu\lambda_- +
{\bar\epsilon}_-\gamma_\mu\lambda_+\ ,  \cr
\delta \phi_i &=-{\bar\epsilon}_-\gamma_i\lambda_-
+{\bar\epsilon}_+\gamma_i\lambda_+\ , \cr
\delta\lambda_+ &= -\ft14 F^{\mu\nu}\gamma_{\mu\nu}\epsilon_+
  -\ft12 \gamma^\mu\gamma^i D_\mu\phi_i \epsilon_-
-\ft14 \gamma^{ij}[\phi_i,\phi_j]\epsilon_+\ , \cr
\delta\lambda_- &= -\ft14 F^{\mu\nu}\gamma_{\mu\nu}\epsilon_-
  +\ft12 \gamma^\mu\gamma^i D_\mu\phi_i \epsilon_+
-\ft14 \gamma^{ij}[\phi_i,\phi_j]\epsilon_-\ , \cr
} \eqno(14)
$$
where $\epsilon$ and $\lambda$ carry the spinor indices of the
 Lorentz group SO(2,2)
as well as the spinor indices of the internal symetry group SO(3,3),
${\bar\epsilon}=
\epsilon^T C_4  C_6$ with $C_4=\gamma_{12}$, $C_6=\gamma_{345}$,
$A_i\equiv \phi_i$,
$\gamma_5 \epsilon_\pm=\pm \epsilon_\pm$ and
$\gamma_7 \epsilon_\pm=\mp
\epsilon_\pm $.  Now it can be seen that $\lambda_+ $ can not be
consistently set equal to zero
in order to arrive at the self-duality equation and its super-partners.
 Unlike in the case of
N=2
super Yang-Mills, here the second and third terms in the transformation
 rule of $\lambda_+ $
can not be made to vanish by any condition on the scalar fields or the
parameters
$\epsilon_\pm$ in a way consistent with supersymmetry [2,8].   A different
kind of self-dual
N=4
super Yang-Mills theory has been formulated in [2], where it is also shown
that it arises as the
effective theory in the open N=2 superstring theory.  In this version, one
 introduces an
anti-self-dual  Lagrange multiplier field  whose equation of motion
 imposes the
self-duality condition on the Yang-Mills curvature, and an anti-chiral
 spinor.  The Lagrange multiplier field propagates, however, and
consequently the number of
helicities in the theory do not get halved.
\bigskip
\noindent{\bf 6.} \  We now turn to the description of two types of N=1
self-dual supergravity
theories in (2,2) dimensions.  One of them, which we call type I, can  be
 obtained  by chiral
truncation of the simple supergravity in (2,2) dimensions.  This truncation
is very much like
the one discussed  above in the case of super Yang-Mills theory [7].
The remaining fields are
$(e_\mu^a, \psi_{\mu -})$,
the vierbein and the gravitino fields, respectively.
The resulting
field equations are
$$
\eqalign{
R_{cd}{}^{ab}(\omega(e)) -
\ft12 \epsilon_{cdef} R^{efab}(\omega(e)) &=0, \cr
\psi^{ab}_- -  \ft12\epsilon^{abcd}\psi_{cd-} &=0, \cr
}\eqno(15)
$$
where  $R_{cd}{}^{ab}$ is the Riemann curvature and
$\psi_{ab}=e^\mu_a e^\nu_b\left[D_\mu(\omega(e)
\psi_\nu-D_\nu(\omega(e)\psi_\mu\right]$.
 The torsion-free spin connection $\omega(e)$ is the solution of the
equation $D_\mu(\omega(e))e_\nu^a\equiv \partial_\mu e_\nu^a +
\omega_\mu{}^{ab} e_{\nu b} -\Gamma^\lambda_{\mu\nu}(g)
e_\lambda^a =0$,  where
$\Gamma^\lambda_{\mu\nu}(g)$  is the usual Christophel symbol.
The solution is given by
$\omega_{\mu ab}=\ft12 e^\nu_a (\partial_\mu e_{\nu b}
-\partial_\nu e_{\mu b})
  +\ft12 e_{\mu c}e^\rho_a e^\sigma_b \partial_\sigma e_\rho^c -
a\leftrightarrow b$.

The type I self-dual supergravity system described in  (15)  has the local
supersymmetry
$$
\eqalign{
\delta e_\mu{}^a &= \ft12\bar\epsilon_+\gamma^a\psi_{\mu-}\ ,\cr
\delta \psi_{\mu-} &= {\cal D}_\mu(\omega(e))\epsilon_- \ .   \cr
}\eqno(16)
$$
Closure of the algebra requires that the supersymmetry parameter
$\epsilon_+$ be covariantly constant, i.e. $D_\mu(\omega)\epsilon_+=0$.
 The integrability
condition of this equation leads to the self-duality condition (15) for
the Riemann curvature.
For later convenience we record here the algebra of the transformations (16):
 $$
[\delta_{\epsilon_1},\delta_{\epsilon_2}] = \delta_\xi+\delta_{\Lambda}
+\delta_{\epsilon_3}\ , \eqno(17)
$$
where the composite general coordinate , Lorenz and supersymmetry
parameters are given by
$$
\eqalign{
\xi^\mu &={\bar\epsilon}_{2+}\gamma^\mu\epsilon_{1-}-
1\leftrightarrow 2\ , \cr
\Lambda &=\xi^\nu\omega_{\nu ab}\ , \cr
\epsilon_3 &= -\xi^\nu\psi_{\nu-}\ .  \cr
}\eqno(18)
$$
In deriving this result, we have used the following formula for the
supersymmetric variation of
the spin connection:
$$
\eqalign{
   \delta\omega_{\mu ab}(e) =& -\ft12 {\bar \epsilon}_+
\gamma_\mu\psi_{ab-}
 +\ft12 {\bar \epsilon}_+\gamma_b\psi_{\mu a-}
 -\ft12 {\bar \epsilon}_+\gamma_a\psi_{\mu b-}  \cr
&\left( \ft12 {\bar \psi}_{\mu -}\gamma_b D_a\epsilon_+
 -\ft12 {\bar \psi}_{a -}\gamma_b D_\mu\epsilon_+
 +\ft12 {\bar \psi}_{b -}\gamma_\mu D_a\epsilon_+  -
a \leftrightarrow b\right) \ . \cr
}\eqno(19)
$$
and the Fierz rearrangement formulae:
$\chi\bar\psi=-\ft12 \bar\psi\chi +
\ft18\bar\psi\gamma^{\mu\nu}\chi\gamma_{\mu\nu}$ for
same chirality anti-commuting Majorana-Weyl spinors and
$\chi\bar\psi=-\ft12
\bar\psi\gamma^\mu\chi\gamma_\mu$ for the case of opposite chiralities.
 A useful
identity is:
$\gamma^{\mu\nu}=-\ft12\epsilon^{\mu\nu\rho\sigma}
\gamma_{\rho\sigma}\gamma_5$.

 Going back to the self-duality equations (15), we note  that there are
no fermionic corrections
to the one involving the Riemann tensor.  Nonetheless one verifies that
these equations do
transform into each other under the supersymmetry  transformations (16),
thanks to the
gravitino equation  $\gamma^{cab}\psi_{ab-}=0$  and its consequences
$\gamma^{ab}
\psi_{ab-}=0$ and $\gamma^a\psi_{ab-}=0$.  The gravitino equation has
to be satisfied by
closure requirements. It can also seen as a consequence of the
self-duality equations (15) taken
together with the Bianchi identities for the curvatures.  This is similar
to the fact that the
self-duality condition on the Yang-Mills field strength plus the Bianchi
 identity imply the
ordinary Yang-Mills field equation.  Notice also that one can derive the
self-duality equation
from the gravitino  equation. To see this we multiply
$\gamma^{cab}\psi_{ab-}=0$  by
$\gamma^d$ and use the equations $\gamma^{ab}\psi_{ab-}=0$ and
$\gamma^a\psi_{ab-}=0$.

Some other aspects of the self-duality condition  (15) are also noteworthy.
{}From (15) and the
identity $R_{[abc]d}(\omega(e))=0$, it  follows that the Ricci-tensor vanishes
$R_{ab}(\omega(e))=0$.
We also know  that Ricci-flat self-dual metrics are Kahler. Therefore, it
follows that self-dual
metrics are  Kahler.  For such metrics, using complex coordinates labeled by
$i=1,2$,  one can
express the Ricci tensor as
$R_{i\bar j}=\partial_i\partial_{\bar j} {\rm ln\ det }g_{i\bar j}$,
where $g_{i\bar j}$ is the metric. In terms of a Kahler potential
$K$, the metric can be
written as $g_{i\bar j}=\partial_i\partial_{\bar j} K$.  Therefore,
the vanishing of the Ricci
tensor means that ${\rm det}\ \partial_i\partial_{\bar j}K=1$. This can
be written as a Poisson bracket as
$\epsilon^{ij}\{\partial_i K,\partial_j K\}=2$, where the
Poisson brackets of $A$ and $B$ is defined as $\{A,B\}=
\epsilon^{\bar i\bar j}\partial_{\bar
i}A\partial_{\bar j}B$. This equation is clearly invariant under
symplectic diffeomorphisms of
the form  $\delta \partial_i K=\{\partial_i K,\Lambda\}$ where $\Lambda$
is an arbitrary
function.  Note that  the  metric is not invariant under this transformation,
and consequently,
from a given metric which satisfies the Ricci flatness condition, one obtains
a one parameter
family of metrics which also obey this condition.
\bigskip
\noindent{\bf 7.}\  We now  describe another type  of N=1
self-dual supergravity theory, which we shall call type II, which can be
obtained from the
so-called new minimal formulation [9] of the off-shell  supergravity
(adapted to (2,2)
signature).  The new minimal multiplet consists of  the vierbein $e_\mu{}^a$,
the gravitini
$\psi_{\mu\pm}$, the antisymmetric tensor field $B_{\mu\nu}$ and
 an SO(1,1) gauge field
$V_\mu$.  It is  to define the combinations
$$
\eqalign{
\Omega_{\mu\pm}{}^{ab} &=
\omega_\mu{}^{ab}(e,\psi) \pm \hat H_{\mu }{}^{ab}\ , \cr
V_{\mu+} &= V_\mu + \ft16\epsilon_\mu{}^{abc}\hat H_{abc}\ , \cr
}\eqno(20)
$$
where the supercovariant spin-connection  $\omega_\mu{}^{ab}(e,\psi)$
 is defined as
$$
\omega_\mu{}^{ab}(e,\psi) = \omega_\mu{}^{ab}(e)
- \ft12 \bar\psi_\mu\gamma^{[a}\psi^{b]} -
\ft14\bar\psi^a\gamma_\mu\psi^b\ , \eqno(21)
$$
with $\psi_\mu=\psi_{\mu+}+\psi_{\mu-}$, and the supercovariant
curvatures are
$$
\eqalign{
\psi_{\mu\nu\pm} &= {\cal D}_\mu(\Omega_+,V_+)\psi_{\nu\pm}
- {\cal D}_\nu(\Omega_+,V_+)\psi_{\mu\pm}\ ,\cr
\hat H_{\mu\nu\rho} &=\partial_{[\mu} B_{\nu\rho]} -
\ft32\bar\psi_{[\mu+}\gamma_\nu\psi_{\rho]-} \ , \cr
\hat F_{ab}(V_+) &=2\partial_{[a}V_{b]+}-
\ft12 \bar\psi_{[a+}\gamma^\lambda
\psi_{b]\lambda-} +
\ft12 \bar\psi_{[a-}\gamma^\lambda\psi_{b]\lambda+}\ ,\cr
\hat R_{cd}{}^{ab}(\Omega_-) &= R_{cd}{}^{ab}(\Omega_-)
-\bar\psi_{[c+}\gamma_{d]}\psi_-^{ab} -
\bar\psi_{[c-}\gamma_{d]}\psi_+^{ab}\ . \cr
}\eqno(22)
$$
The derivative ${\cal D}_\mu$ on the supersymmetry parameters
$\epsilon_\pm$ is given by
 $$
{\cal D}_\mu(\Omega_+,V_+)\epsilon_\pm = \biggl (
\partial_\mu - \ft14 \Omega_{\mu+}{}^{ab}\gamma_{ab} \mp
V_{\mu+}\biggr ) \epsilon_\pm \eqno(23)
$$

The N=1 type II self-dual supergravity system consists of the
 following equations [5]
$$
\eqalign{
\psi^{ab}_\pm  &=  \ft12\epsilon^{abcd} \psi_{cd\pm}\ ,   \cr
\hat R_{cd}{}^{ab}(\Omega_-) &=
\ft12 \epsilon^{abef} \hat R_{cdef}(\Omega_-)\ ,  \cr
\hat F^{ab}(V_+) &= \ft12\epsilon^{abcd}\hat F_{cd}(V_+)\ .  \cr
}\eqno(24)
$$
These equations transform into each other under the following local
supersymmetry
transformations
$$
\eqalign{
\delta e_\mu{}^a &= \ft12 \bar\epsilon_+\gamma^a\psi_{\mu-}+
\ft12\bar\epsilon_-\gamma^a\psi_{\mu+}\cr
\delta\psi_{\mu\pm} &= (\partial_\mu -
\ft14\Omega_{\mu+}{}^{ab}\gamma_{ab}
\mp V_{\mu+})\epsilon_\pm \cr
\delta B_{\mu\nu} &= \ft32\bar\epsilon_+
\gamma_{[\mu}\psi_{\nu]-}
 + \ft32\bar\epsilon_-\gamma_{[\mu}\psi_{\nu]+}\cr
\delta V_\mu &= \ft18\bar\epsilon_+\gamma_\mu\gamma^{ab}
\psi_{ab-} \cr
}\eqno(25)
$$

It is interesting to observe [5] that the last self-duality condition in
(24) together with the
Bianchi identity $D_{[a}\hat F_{bc]}(V_+)+\cdots=0$ imply the field
equation $D_a \hat
F^{ab}(V_+) +\cdots=0$, which we recognize as one of the field
equations that follow from the
(2,2) version of the $R^2$-type action constructed in [10] given by
$$
\eqalign{
e^{-1}{\cal L}(R^2) =& \ft14 R_{\mu\nu}{}^{ab}(\Omega_-)
R^{\mu\nu ab}(\Omega_-) - 2 \hat F_{ab}(V_+)\hat F^{ab}(V_+)
+ \ft12\bar\psi^{ab}\gamma^\mu D_\mu(\Omega_-,V)\psi_{ab} \cr
&+\ft18\bar\psi_\mu\gamma^{cd}\gamma^\mu\psi_{ab} \bigl\{
R_{cd}{}^{ab}(\Omega_-)+\hat R_{cd}{}^{ab}(\Omega_-)\bigr\}  \cr}
\eqno(26)
$$
where $\psi_\mu=\psi_{\mu+}+\psi_{\mu-}$ and the $\Omega_-$
covariantization
in the kinetic term of the gravitino curvature acts both on the spinor
as well as the vector indices of $\psi^{ab}$.
By supersymmetry it then follows that the field equations for the remaining
fields $e_\mu{}^a, \psi_{\mu\pm}$ and $B_{\mu\nu}$ must also be satisfied.

The N=2 and N=4 self-dual supergravity theories in (2,2) dimensions can also be
 obtained
from their  non-self-dual counterparts (which in turn can be  obtained
by a Wick
rotation of
their (3,1) versions) by  chirality/self-duality restrictions [8,2].
The case of N=8
supergravity, just as in the case of N=4 super Yang-Mills, turns out to
be more subtle, and
has been treated in [2].
 \bigskip
\noindent{\bf 8.}\
In [2], it was found that a self-dual version of {\it gauged} N=8 supergravity
arsies as an effective theory of closed N=2 superstrings. The relevant global
superalgebra, in the case of non-self dual extended supergravity in (2,2)
 dimensions,
is  $OSp(N|4)$, whose bosonic subalgebra is $SO(3,2)\oplus SO(n,N-n)$.
It can be
obtained from the superconformal algebra $SL(N|4)$, which has the
 bosonic subalgebra
$SO(3,3)\oplus SO(1,1)\oplus SL(N,R)$, by a suitable restriction.  In the
resulting algebra, the
SO(3,2) generators  can be decomposed with respect to the SO(2,2)
 subalgebra as $(M_{\mu\nu},
P_\mu)$ and in addition to the obvious commutation rules one has
 $$
 \eqalign{
\{Q_+^i, Q_+^j\} &= \gamma^{\mu\nu}C  M_{\mu\nu}^-\eta^{ij}
+\ft12 (1+\gamma_5)C T^{ij}\ , \cr
\{Q_+^i, Q_-^j\} &= \ft12 (1+\gamma_5)\gamma^\mu C  P_\mu\eta^{ij}
\ ,  \cr
\{Q_-^i, Q_-^j\} &= \gamma^{\mu\nu}C  M_{\mu\nu}^+\eta^{ij}
+\ft12 (1-\gamma_5)C T^{ij}\ , \cr
[P_\mu, Q_\pm^i] &= \ft12\gamma_\mu Q_\mp^i\ , \qquad\qquad
[M_{\mu\nu}^{\mp}, Q_{\pm}]=\ft12 \gamma_{\mu\nu} Q_{\pm}\ ,    \cr
}\eqno(27)
$$
where $\eta^{ij}$ is the invariant tensor, and $T^{ij}$ are the generators,
of $SO(n,N-n)$, and
$M_{\mu\nu}^{\pm}$ are the self-dual and anti-self dual pieces of
the Lorentz generators.

The  gauged N=8 supergravity  in (2,2) dimension  is rather complicated.
Its self-dual
restriction is in some ways similar to the case of N=4 super Yang-Mills
theory, where
propagating Lagrange multipliers are needed to impose the self-duality
condition.  An
important property of gauged supergravities is that they contain a cosmological
constant.
While the cosmological constant is necessary in  gauged supergravity theories,
gauging is not
necessary in order to introduce a cosmological constant.  The simplest
example in four dimensions is  N=1 supergravity  with cosmological
constant [11].  It is
instructive to consider the consequences of its Wick rotation to (2,2)
dimensions and its chiral
truncation.  The transformation rules in (2,2) dimensions are
$$ \eqalign{
\delta e_\mu^a &={\bar\epsilon}_+\gamma^a\psi_{\mu -}
+{\bar\epsilon}_-\gamma^a\psi_{\mu +}\ ,  \cr
\delta\psi_{\mu +} &=D_\mu({\hat\omega})\epsilon_+ +
m\gamma_\mu\epsilon_-\ , \cr
\delta\psi_{\mu -} &=D_\mu({\hat\omega})\epsilon_- +
m\gamma_\mu\epsilon_+\ , \cr
}\eqno(28)
$$
where $m$ is an arbitrary mass parameter, and
${\hat\omega}_{\mu ab}$  equals
$\omega(e)_{\mu ab}$
plus fermionic torsion which consists  of bilinears in the gravitino field.
The closure of this
algebra requires the field equations  that follow from the  supergravity
action plus a gravitino
mass term and a cosmological constant term.  In particular,
$\gamma^{\mu\nu\rho}\psi_{\nu\rho -}
\sim  m\gamma^{\mu\nu}\psi_{\nu +}$ and
$\gamma^{\mu\nu\rho}\psi_{\nu\rho +}
\sim  m\gamma^{\mu\nu}\psi_{\nu -}$.

  Now let us consider  the  chiral truncation $\psi_+=0$.
{}From (28) we get the condition $D_\mu (\omega(e))\epsilon_+ +
m\gamma_\mu\epsilon_-=0$,
which determines
$\epsilon_-$ in terms of $\epsilon_+$. However, we run into a
problem with the  gravitino
field equation: $\gamma^{\mu\nu}\psi_{\nu-}=0$, and hence
$\psi_{\mu -}=0$.  Interestingly enough, there is another route
one can take, which
turns out to imply that the torsion-free  self-dual N=1 supergravity
equations actually admit
a one parameter extension of the super Poincar\'e algebra as a
symmetry algebra.  To see this,
let us consider the folowing transformations
$$
\eqalign{
\delta e_\mu^a &={\bar\epsilon}_+\gamma^a\psi_{\mu -}  \ ,  \cr
\delta\psi_{\mu-}&=D_\mu(\omega(e))\epsilon_- +
m\gamma_\mu\epsilon_+\ . \cr
}\eqno(29)
$$
Remarkably,  these transformation rules do form a closed
algebra provided that
$D_\mu(\omega(e))\epsilon_+=0$ and the
self-dual supergravity equations (15) are satisfied.  The algebra of
these transformation is
$$
[\delta_{\epsilon_1},\delta_{\epsilon_2}] = \delta_\xi+\delta_{\Lambda'}
+\delta_{\epsilon_3}\ , \eqno(30)
$$
where the composite general coordinate, Lorenz and supersymmetry
parameters are given by
$$
\eqalign{
\xi^\mu &={\bar\epsilon}_{2+}\gamma^\mu\epsilon_{1-}-
1\leftrightarrow 2\ , \cr
\Lambda' &=\xi^\nu\omega_{\nu ab}+2m{\bar\epsilon}_{2+}
\gamma_{ab}\epsilon_{1+}\ ,  \cr
\epsilon_3 &= -\xi^\nu\psi_{\nu-}\ .  \cr
}\eqno(31)
$$
One can check that $\delta_{\epsilon_3}e_\mu^a=0$ and that
$\delta_{\Lambda'}\psi_{\mu -}=\delta_{\Lambda}\psi_{\mu -}$.  We see
 that unlike in the case
of the algebra (17), here even in the global limit  we still have an (anti)
self-dual   Lorentz
rotation.  Noting that $[\delta_\epsilon,\delta_\xi]=\delta_{\epsilon'}$
and $[\delta_\epsilon,
\delta_\Lambda]=\delta_{\epsilon''}$ where
$\epsilon'=-\xi^\mu\partial_\mu \epsilon$ and
$\epsilon''=-\ft14 \Lambda_{ab}\gamma^{ab}\epsilon$, we see that
the global limit of the full
algebra contains the generators $(Q_\pm, M_{\mu\nu}^-, P_\mu)$,
 obeying the algebra
$ \{Q_+, Q_+\}= \gamma^{\mu\nu}C  M_{\mu\nu}^-$,
$\{Q_+, Q_-\} = \ft12 (1+\gamma_5)\gamma^\mu C  P_\mu$,
and $[M_{\mu\nu}^-, Q_+]=\ft12 \gamma_{\mu\nu} Q_+$, plus  the
commutators of the ordinary
Poincar\'e algeba, with all the other (anti) commutators vanishing.

\bigskip\bigskip

\centerline{\bf Acknowledments}

I thank Mike Duff, Chris Pope and Larry Romans for stimulating discussions.

\vskip 2truecm
\centerline{\bf References}
 \bigskip
\item{1.}
H.\ Ooguri and C.\ Vafa,  Mod.\ Phys.\ Lett.  {\bf A5} (1990)
1389;  Nucl.\ Phys. {\bf B361} (1991) 469;  Nucl.\ Phys. {\bf B367} (1991)
83.
\item{2.}
W.\ Siegel,  preprints, ITP-SB-92-24, 31,53.
\item{3.}
See, for example, the review article by:
\item{}  R.S.\ Ward in {\it Field Theory, Quantum Gravity and
Strings}, Vol.\ 2, p.\ 106, eds.\ H.J.\ De Vega and N.\ Sanchez.
\item{4.}
 S.J.\ Gates Jr., H.\ Nishino and S.V.\ Ketov, preprint, UMDEPP 92-51.
\item{5.}
E. Bergshoeff and E. Sezgin, preprint, UG-4/92, CTP TAMU-46/92.
\item{6.} For the properties of spinors in arbitrary dimensions, see:
\item{} T.\ Kugo and P.K.\ Townsend, {\it Nucl.\ Phys.} {\bf B221}
(1983) 357.
\item{7.}
S.M.\ Christensen and M.J.\ Duff,  Nucl. Phys.  {\bf B159}(1979) 429;
\item{}
S.M.\ Christensen, S.\ Deser,\ M.J.\ Duff and M.T.\ Grisaru,  Phys. Lett.
{\bf 84B}
(1979) 411;
R.E.\ Kallosh,  JETP Lett.  {\bf 30} (1979)172,449;  Nucl. Phys.
{\bf B165} (1980)
 119.
\item{8.}
 S.J.\ Gates Jr., H.\ Nishino and S.V.\ Ketov, preprints,
UMDEPP 92-163,171,187,211.
\item{9.}
M.F.\ Sohnius and P.C.\ West,  Phys.\ Lett. {\bf 105B} (1981) 353.
\item{10.}
M.\ de Roo, A.\ Wiedemann and E.\ Zijlstra,  Class.\ Quantum Grav.
{\bf 7} (1990) 1181.
\item{11.}
 P.K. Townsend,  Phys. Rev. {\bf D15} (1977) 2808.

\end